\documentclass[pra,twocolumn]{revtex4-1} 

\usepackage{amsmath}  
\usepackage{amsfonts} 
\usepackage{graphicx} 

\newcommand{\be}{\begin{equation}}
\newcommand{\ee}{\end{equation}}
\newcommand{\ba}{\begin{align}}
\newcommand{\ea}{\end{align}}

\newcommand\dip[1]{\mathbf{d}_{#1}(\mathbf{k}_t)}

\newcommand\den[1]{\rho_{#1}(\mathbf{k};t)}

\newcommand\eps[1]{\epsilon_{#1}(\mathbf{k}_t)}

\newcommand\kk{\mathbf{k}}
\newcommand\kt{\mathbf{k}_t}

\begin{document}

\title{On the (un)importance of the transition-dipole phase in the high-harmonic generation from solid state media}

\author{J.~Gu}
\author{M.~Kolesik}
\affiliation{James Wyant College of Optical Sciences, The University of Arizona, Tucson, AZ 85721, U.S.A. }

\begin{abstract}
  Solid-state high-harmonic generation (HHG) continues to attract a lot of interest.
  From the theory and simulation standpoint, two issues are still open;
  The first is the so-called transition-dipole phase problem. It has been recognized that the dipoles must
  be treated as complex-valued quantities, and that their corresponding Berry connections must be included
  to ensure phase-gauge invariance. However, while this has been successfully implemented for lower-dimensional systems,
  fully vectorial and three-dimensional simulations remain to be challenging.
  The second issue concerns the symmetry of the high-harmonic response, when simulations sometimes
  fail to honor the symmetry of the crystalline material.
  This work addresses both of these problems with the help of a HHG-simulation
  approach which 
  a) is manifestly free of the transition-dipole phase problem,
  b) does not require calculation of dipole moments,
  c) can account for the contributions from the entire Brillouin zone,
  d) faithfully preserves the symmetry of the simulated crystalline material.
  We use the method to show that high-harmonic sources are distributed throughout the Brillouin zone
  with various phase-shifts giving rise
  to significant cancellations. As a consequence, for the simulated response to correctly capture
  the material symmetry, contributions from the entire Brillouin zone must be included. Our results
  have important implications for a number of HHG applications, including all-optical band- and dipole-reconstruction.
\end{abstract}
    
\maketitle

\section{Introduction}

High-harmonic generation in solid-state media has been studied with keen interest
ever since the first observations a decade ago~\cite{Ghimire11} followed by
experiments with many different materials and structures~\cite{BrabecRev,recentHHGtrends}.
Mediated by the light-matter interactions at high density, the phenomenon opens
a new window into the dynamics of  the solid-state medium at attosecond time-scales,
including
all-optical reconstruction of the  band structure~\cite{ReconstAllOptBands,ReconstBands,MapAssume0},
mapping of the transition-dipole moments~\cite{ReconstDipole},
characterization of higher-order nonlinearity~\cite{ExtractHOKE},
and measurements of Berry curvatures~\cite{BerryCurvature}.

Numerical simulations have played an important role in this field~\cite{BrabecRev,HHGmethods,Gaarde22}.
The  broad spectrum of applied approaches ranges from the ab initio time-domain Schr\"odinger
equations~\cite{TDSEPlaja},  multiscale time-domain density-functional theory~\cite{Multiscale},
through many variants of the semiconductor Bloch equations (SBE) and density-matrix methods~\cite{KochSBE,KochBook,KiraBook,Gaarde22},
to the studies including propagation effects~\cite{Xia:18} and coupling with Maxwell equations~\cite{Multiscale,Kilen20,FullBZMaxwel,hhgnlo}.

One of the issues that attracted attention over the last few years is that of the transition-dipoles.
It concerns the phase-gauge degree of freedom in the description of the electronic Bloch states; they can
be modified by arbitrary phase factors~\cite{BrabecRev,Gaarde22,Lindefelt} which in turn change the complex phase of the off-diagonal
dipole-matrix elements. Closely related to this is the Berry connection which, assuming that Bloch-states
are differentiable, gives a gauge-dependent measure of how the Bloch basis changes from
one point to the next over the Brillouin zone.

While semiconductor Bloch equations are  phase-gauge invariant~\cite{phsinv}, some early simulations broke this
symmetry with the dipole moments treated as real-valued quantities (see discussions in~\cite{TDP1,TDP2,TDP3}).
Moreover, Berry connections~\cite{BrabecRev,Gaarde22} are still often neglected,
which also breaks the gauge-invariance of SBEs.
The proper treatment requires the construction of a differentiable~\cite{smoothphs}
and Brillouin-zone periodic phase-gauge~\cite{smoothTDP}.
Imposing such a phase-gauge have been demonstrated in one-dimensional models, but
doing the same in three-dimensional reciprocal space have not been shown explicitly yet.
At any rate, the fact that the Berry connections and dipole-moment phases
need careful attention makes the simulation of the HHG from crystals even more difficult --- this is what we refer to
as the transition-dipole phase (TDP) problem.

Another issue complicating the modeling is that in principle all states from the  Brillouin zone 
contribute to HHG. Currently only a few approaches account for the full three-dimensional Brillouin zone (e.g.~\cite{Multiscale,FullBZMaxwel,TDSEPlaja,fullGaSe}),
and this requires
extreme computational efforts. In contrast,  most of the modeling to date has been done with lower-dimensional
spaces such as straight paths across the center of the Brillouin zone, raising a question if
the chosen subset really dominates the HHG process~\cite{Imasaka:22}. Efficient methods which include all
Bloch states are therefore needed.

Intimately related to these two problems is the issue of the symmetry. Clearly, at least for the low excitation
intensity the simulated medium response must have the symmetry dictated by the space group of the crystal. For example,
the simulated second-order nonlinear tensor must exhibit ``hard zeros'' where the symmetry implies vanishing components.
This sometimes proved problematic (see e.g. Refs.\cite{TDP2,TDP3} for a discussion and references therein),
when earlier simulations failed to produce and/or to suppress even harmonics as required by the symmetry of the
problem.

One of the goals of this work is to put forward a HHG-simulation approach which addresses these issues.
It is designed with the recognition that the phase-problem is very much self-imposed, it is in fact not required
by physics and can be by-passed~\cite{IvanovWanier}. By eliminating any and all phase-gauge dependencies, the resulting method
i) is manifestly free of the TDP-problem because it does not require dipole moments or the Berry connection in the first place,
ii) it can efficiently add-up the HHG-contributions from the full Brillouin zone,
and iii) it automatically produces the response with the correct symmetry.
It should be emphasized that the algorithm, while oblivious to Berry connections and to
transition dipoles, does not neglect them. Instead, it can work with arbitrary phases
implicitly assigned to the band-structure states. In this respect, the approach is distinctly
different from other treatments of the TDP issue, including the Wannier representation~\cite{IvanovWanier}. 

We utilize this tool to gain insight into how the HHG is sourced across the Brillouin zone.
We show that significant destructive interferences can occur between the HHG contributions originating from
distant parts of the Brillouin zone. Moreover, it is not always the case that  the regions with the
strongest dipoles dominate the generated radiation. These observations imply that simplified models based on
one-dimensional subsets of the reciprocal space must be treated with extreme caution, while the full-3D approach
should be preferred whenever computationally feasible.

\section{Semiconductor Bloch Equations}

Semiconductor Bloch Equations~\cite{KochSBE,KochBook,KiraBook} represent one of the most frequently utilized approaches to
the high-harmonic generation in solid-state media~\cite{Gaarde22}. For the sake of completeness,
we review the most important components of the method in this section. We choose to follow
Ref.\cite{Wilhelm21} by Wilhelm et al., and refer the reader to this well-rounded exposition for details.

It is assumed for this work that the excitation by an optical pulse is at mid-infrared
or longer wavelength for which the interaction with the material can be considered off-resonance.
Consequently, the Coulomb interactions play a lesser role~\cite{KochTe} and are neglected in what follows.
Note that this may not be justified for effectively two-dimensional materials~\cite{HHG2D},
but HHG from bulk crystals is  often treated this way.

Assuming that the band-structure of the material is known throughout the  Brillouin zone,
let $\epsilon_n(\mathbf{k})$ with $n=1,\ldots, N_b$ describe the $N_b$ energy bands with
corresponding eigenvectors $\{\vert n \mathbf{k}\}\rangle$.
The quantum state of the system is given by the density matrix $\den{mn}$ with $\mathbf{k}$ running
over the Brillouin zone.
The initial condition before the excitation pulse
arrives is approximated by the zero-temperature density matrix with all
conduction bands completely empty and valence bands full.

\smallskip

\noindent {\bf Evolution equations for the density matrix:}\\
The SBE system constitutes a set of coupled differential equations, which can be represented
in a number of equivalent ways and gauges (described in a recent tutorial by Yue and Gaarde~\cite{Gaarde22}).
Here it is written in the time-dependent basis  $\{\vert n \mathbf{k}_t\rangle\}_n$
as an evolution equation for the density matrix
$\den{nm}$, 
\begin{align}
\label{eq:sbe}
  (i\partial_t - \eps{nm})&\den{nm} = \\
\mathbf{E}(t)\sum_a &\left(\den{na}\dip{am} -\dip{na} \den{am}\right) \nonumber 
\end{align}
where the dipole-moment matrix
\be
\dip{am} = \langle a \mathbf{k}_t\vert i \partial_{\mathbf{k}_t}\vert m \mathbf{k}_t\rangle
\label{eq:dipole}
\ee
and the band-energy differences 
\be
\eps{nm} = \eps{n}\!-\!\eps{m}
\ee
are calculated for the time-dependent k-vector
\be
\mathbf{k}_t = \mathbf{k} - \mathbf{A}(t) \ 
\label{eq:kshft}
\ee
reflecting the effect of the electromagnetic vector potential $\mathbf{A}(t)$ of the excitation pulse.
For the moment, de-phasing terms are omitted for the sake of simplicity --- they will be included
later.

Equations (\ref{eq:sbe}) to (\ref{eq:kshft}) are in the velocity gauge. One advantage over their
counterpart in the length gauge is that the latter contains gradients which result in a coupling between
equations for different $\mathbf{k}$. This version is therefore easier to parallelize with a near-perfect
load balance. Because we integrate the evolution for all relevant Bloch states, the parallel efficiency
is an important aspect to consider.

\smallskip
\noindent{\bf Observables:}
Once the evolution system is integrated for all $\mathbf{k}$, the induced current density
is calculated by integrating
the  Brillouin zone and adding contributions from all bands (formula (62) in Ref.~\cite{Wilhelm21}) like so
\be
\mathbf{j}(t) =
\sum_{mn}\int\frac{d\kk}{(2\pi)^3} \langle n\kt\vert \partial_{\kt} h(\kt)\vert m \kt\rangle \rho_{mn}(\kk ;t) \ .
\label{eq:observeJ}
\ee
Here, $h(\mathbf{k})$ is the instantaneous Hamiltonian with eigenstates $\{\vert n \mathbf{k}\rangle\}$
corresponding to the given k-vector, and $\partial_{\mathbf{k}} h(\mathbf{k})$ is the Hamiltonian-matrix gradient
in the reciprocal space.
Note that the current density can be separated into various components~\cite{Wilhelm21},
including inter- and intra-band contributions~\cite{InterIntra} for more physical insight, but this is not pursued here.


With the current density coupled to Maxwell equations, ``all one needs to do'' to simulate high-harmonic
generation in a medium exposed to an electromagnetic pulse is to integrate the Maxwell-SBE system.
However, the above equations were derived with certain assumptions which bring complications. 
One has to evaluate the dipole moment operator (\ref{eq:dipole}) which obviously requires
$\vert m \mathbf{k}\rangle$ to be differentiable with respect to $\mathbf{k}$. This is where the
transition-dipole phase issue comes in.

\section{Transition-dipole phase}

There is an extensive literature dealing with the so-called transition-dipole phase (TDP)
problem (see e.g.~\cite{smoothTDP}). Not a long time ago, in the early simulations of HHG from solids, the fact that
the dipole-moment as a function of the k-vector is a complex-valued quantity was 
ignored and only the absolute values were utilized in the calculations. The state of the art
improved in the recent years, and the community has a good understanding of these issues~\cite{Gaarde22}.
Nowadays there is a consensus that the ``transition-dipole phase plays a role'' in HHG
(see e.g. \cite{TDP1,TDP2,TDP3}), but we feel it is useful to
emphasize that the {\em absolute phase} of any dipole matrix element is not a measurable quantity.
This is why we want to include a very brief review here.

\smallskip

\noindent{\bf Gauge invariance:}\\
Let us start with the origin of the TDP-problem. In quantum theory, the state of a system is
represented not by a vector, as it is often inaccurately described in the physics literature, but by
a ray which is a one-dimensional subspace of the Hilbert space (e.g.~\cite{QMpostulatesPRL,QMpostulatesNatComm}).
In other words, after multiplication by an arbitrary non-zero complex number, the vector
still stands for the exactly same physical state.  This means that as $\mathbf{k}$ runs over the
Brillouin zone, bases  $\{\vert n \mathbf{k}\rangle\}$ can be replaced
by ones which differ by arbitrary phase factors on each of their elements, $\{e^{i\phi_n(\mathbf{k})}\vert n \mathbf{k}\rangle\}$,
where the phase $\phi_n(\mathbf{k})$ can be anything, including non-differentiable, non-continuous or
even completely random.

Any change in the chosen phase of the basis vectors by  $\phi_n(\mathbf{k})$ modifies the transition dipole (\ref{eq:dipole})
\be
\mathbf{d}_{am}(\mathbf{k}) \to  e^{-i\phi_a(\mathbf{k})} \langle a \mathbf{k}\vert i \partial_{\mathbf{k}}\vert m \mathbf{k}\rangle e^{+i\phi_m(\mathbf{k})}
\label{eq:dipphase}
\ee
which makes it evident that the SBE system in fact {\em assumes} that the phases of the basis vectors
throughout the Brillouin zone were {\em chosen} such that the resulting dipole moments are differentiable.
This has been called differentiable gauge, and one usually adds a requirement
that the dipoles are also made Brillouin-zone periodic.

Of course, changing the gauge also modifies the off-diagonal elements of the density matrix.
However, once the physical observables are calculated as e.g. in (\ref{eq:observeJ}) the choice of the
phases gets completely ``erased.'' This is a manifestation of the phase-invariance of the SBE system
which has been shown via explicit calculations for various SBE-representations~\cite{phsinv}. The same
conclusion can be obtained already from the basic principles of the quantum mechanics.
Indeed, since the phase-modified basis vectors represent the same physical states, observable quantities are always
completely independent of how  $\phi_n(\mathbf{k})$ may be set. Thus, there is no
measurement which could reveal the absolute phase of a vector or of a  matrix element, including that of
the dipole moment (\ref{eq:dipole}). This does not mean that the dipole moment phase can be set
arbitrarily because one only has $N_b-1$ free parameters to adjust phases of $N_b (N_b-1)/2$ off-diagonal
elements of $\mathbf{d}_{am}(\mathbf{k})$.

\smallskip

\noindent{\bf Numerical issues:}\\
Before running a HHG simulation based on the SBE, one must obtain the dipole
moments. Density functional theory softwares are most often used to calculate the band structure of
a material and they can also provide the dipole matrices. No matter what kind of a solver
is used to diagonalize the model Hamiltonian, the resulting eigenstates calculated
for two nearby k-vectors may or may not end up close to each other. In particular, the
phases of the bases obtained at different location inside Brillouin zone may appear
``random'' (although in practice they are not truly random). For this reason, algorithms to
generate a ``smooth periodic phase'' have been developed~\cite{smoothTDP}. It is relatively straightforward to
obtain a smooth phase along a one-dimensional subspace of the Brillouin zone,
and it can also be arranged to have a desired periodicity. However, to the best of our
knowledge the methods were not yet explicitly demonstrated for the three-dimensional
reciprocal space.

Another consideration relevant for the numerical treatment is the calculation of
the off-diagonal dipole moments and of the Berry connection which is the diagonal part
of $\mathbf{d}_{am}(\mathbf{k})$. The off-diagonal part can be obtained without numerical differentiation~\cite{Wilhelm21},
but this depends on expressions which become numerically inaccurate when close to degeneracy.
Nevertheless, since it is possible to avoid numerical differentiation for the off-diagonal dipoles,
one may wonder if the SBE representation (\ref{eq:sbe}), which does not feature any gradients, needs to
care about the dipole phase at all; is it perhaps possible to execute the simulation
with whatever phases were given to the dipoles by the eigensolver? The answer
would be affirmative if not for two serious issues: i) extremely poor accuracy around
sharp ``phase jumps'' (which are guaranteed to occur) and, more importantly,
ii) the diagonal part, i.e. Berry connection which is a gauge-dependent quantity.

The inclusion of the Berry connection is crucial for maintaining the phase invariance of the system~\cite{phsinv}.
One reason it was possible to ignore it in many simulations is that leaving out Berry connection may
still produce a reasonably looking high harmonic spectrum. Nevertheless, such results are incorrect
because they depend on the nonphysical (as in un-observable) phase choice for the Hamiltonian bases. 
Numerical evaluation of Berry connections  involves ``comparison'' of Hamiltonian bases
at nearby k-vectors. This calculation is essentially similar to numerical differentiation
and it may require an extremely fine grid in the k-vector space.

To summarize this section, once we have committed to simulate
the SBE-system (\ref{eq:sbe}) or  its gauge-related counterparts (see \cite{Wilhelm21}) in the precise
form as written, we must address the problem
of the smooth, Brillouin-zone periodic phases assigned to the states of the material
band-structure. Moreover, we need to evaluate the transition dipole matrix elements
and the Berry connection
which brings a set of further numerical challenges. This begs the question if
all of this is really necessary, because the requirement of the differentiable
TDP is ``self-inflicted'' by the choice of assumptions underlying (\ref{eq:sbe}).
Quantum theory says that for any observable quantity {\em all phase choices are equivalent},
so one could design the SBE-solver to be ``phase-choice oblivious.''
This is demonstrated next.

\section{SBE solver algorithm}

In order to lay out the idea of the algorithm, it should be useful to
appreciate the roles played by the different terms in the SBE-system (\ref{eq:sbe}).
Detailed derivations, as shown e.g. in Refs.~\cite{Wilhelm21,Gaarde22}, make it evident
that the part proportional to the electric field originates from the
time-dependent basis. Even a constant solution appears to depend
on time when a time-dependent basis is used, and it is this what the term
accomplishes upon integration. Namely, it smoothly transforms the density matrix from
the Hamiltonian basis at time $t_1$ to a different basis at time $t_2$. So if
it is sufficient to know the solution  only at these discrete points in time,
we can transform the density matrix with a unitary matrix in a single step,
and thus skip all the work  needed to solve the system of ordinary differential
equations, and avoid accumulation of numerical errors at the same time.

To demonstrate that we get the correct solution, consider the right-hand-side
of (\ref{eq:sbe}) between times $t_i$ and $t_{i+1}$, and construct the following unitary
  matrix
  \be
  U_{ab}(t) = \langle a  \mathbf{k}_{t}\vert b \mathbf{k}_{i}\rangle ,
  \ee
  with
  \be
  \mathbf{k}_t = \mathbf{k} - \mathbf{A}(t) \ \ \text{and} \ \
  \  \mathbf{k}_i = \mathbf{k} - \mathbf{A}(t_i) \ .
  \label{eq:U}
  \ee
  Next, calculate
  \be
  \rho(t) =    U(t)  \rho(t_i)  (U(t) )^\dagger
  \label{eq:splitU}
  \ee
  for $t_i < t < t_{i+1}$, and differentiate it with respect to $t$ to obtain,
  \be
  i \dot \rho(t) =    i \dot U(t)  \rho(t_i)  (U(t) )^\dagger + i  U(t)  \rho(t_i)  (\dot U(t) )^\dagger \ .
  \ee
  Inserting  $I = U U^\dagger = U^\dagger U$
  between the constant $\rho(t_i)$ and the dotted (time-differentiated) operators  
and subsequently  using (\ref{eq:splitU}) we get
  \be
  i \dot\rho(t) =   i \dot U(t) (U(t))^\dagger  \rho(t)   +  i \rho(t)  U(t)  (\dot U(t))^\dagger \ .
  \ee
  Using  $\dot U U^\dagger = - U \dot U^\dagger$ one obtains the right-hand-side in the form
  of a commutator,
  \be
  i \dot\rho(t) =   i \dot U(t) (U(t))^\dagger  \rho(t)   -  i \rho(t) \dot U(t)  (U(t))^\dagger \ ,
  \label{eq:aux}
  \ee
  which is to be compared to that in (\ref{eq:sbe}), so we want to expand $ \dot U U^\dagger $.
  The time derivative of the transformation matrix is
  \be
  \dot U_{ab}(t) = \partial_t \langle a  \mathbf{k}_{t}\vert b \mathbf{k}_{i}\rangle =
 \mathbf{E}(t) . \langle \partial_{\mathbf{k}_t} a  \mathbf{k}_{t}\vert b \mathbf{k}_{i}\rangle \ ,
 \ee
and $\dot U U^\dagger$ reveals the dipole moment and the electric field,
  \begin{align}
  i \dot U_{ab} (U^\dagger)_{bc} = 
  i & \mathbf{E}(t) .  \langle \partial_{\mathbf{k}_t} a  \mathbf{k}_{t}
\vert b  \mathbf{k}_{i} \rangle\langle b  \mathbf{k}_{i}
  \vert c \mathbf{k}_{t}\rangle =  \\
 -&\mathbf{E}(t) . \langle a  \mathbf{k}_{t}\vert i \partial_{\mathbf{k}_t} c \mathbf{k}_{t}\rangle =
 - \mathbf{E}(t) . \dip{ac} \ . \nonumber
 \end{align}
 Using this in (\ref{eq:aux}) gives
  \be
  i\dot\rho_{nm} = 
\mathbf{E}(t)\sum_a \left(\rho_{na}\dip{am} -\dip{na} \rho_{am}\right) 
\ee
which is precisely the $\mathbf{E}$-field term in (\ref{eq:sbe}). Thus, the basis-transformation
(\ref{eq:U}) would give an exact solution if not for the diagonal part of the
SBE system. Because the exact solution can be also obtained for the diagonal part,
Eqn.~(\ref{eq:sbe}) is a natural candidate for the operator-splitting approach.

Let us assume that the evolution of the system is sampled on a discrete grid of times, $t_i$,
and let $\mathbf{k}_i$ stands for  $\mathbf{k}_t$ calculated for $t=t_i$. Further, let
$\{\vert m \mathbf{k}_i\rangle\}_{m=1}^{N_b}$ be the Hamiltonian eigen-basis at time $t_i$,
and we use it with whatever phases an eigen-system solver assigned to the eigenvectors.
The basis transformation between $t_i\to t_{i+1}$ is given by the unitary matrix
\be
U^{(i)}_{ab} = \langle a  \mathbf{k}_{i+1}\vert b \mathbf{k}_{i}\rangle ,
\ee
and this is used as in (\ref{eq:splitU}) to evolve the density matrix
from $t_i$ to $t_{i+1}$.

The other split-operator is diagonal; it represents the adiabatic evolution
in the time-dependent basis. Joining the two split-operator actions together,
the density-matrix evolution over the time-step interval $\Delta t = t_{i+1}-t_i$ can be approximated 
by
\be
\rho(t_{i+1}) = P^{(i)} \rho(t_{i}) (P^{(i)})^\dagger \ ,
\ee
where the evolution operator is
\be
%
P^{(i)}_{ab} = e^{-i \epsilon_a(\mathbf{k}_i) \Delta t/2 } U^{(i)}_{ab} e^{-i \epsilon_a(\mathbf{k}_i) \Delta t/2 } \  .
\ee
This operator-splitting formula is locally second-order accurate~\cite{MSuzuki90}, but that alone
tells us little about how long $\Delta t $ can be.
At any rate, the permissible integration step depends on the electric field intensity
and it must be established in a case-by-case convergence study. For the simulation results
presented in this work the time step was $\Delta t=0.07$~fs. 

In order to complete one integration step, the phenomenological damping can be included between
the split-steps  by appropriate modification of the off-diagonal parts of the density matrix~\cite{Gaarde22}.
We used a dephasing time of five femtoseconds for our examples in Section VI.

Since we have not assumed anything about the phase-relations between the bases at $t_i$ and $t_{i+1}$,
the Hamiltonian eigenstates can be used as calculated by the eigensystem solver, and this algorithm is manifestly
free of the transition-dipole phase problem. As a sanity check, we have inserted in the numerical evolution scheme
a procedure which generates and assigns truly random phases to all Hamiltonian eigenvectors after each and every call to the eigensystem solver --- with no significant change in observables.

The fact that we can work with {\em any} eigenvector phases including random ones is a crucial advantage over the approaches which
rely on the numerical integration of SBE using ODE-solvers. An additional important benefit
is that this algorithm does not require calculation of the dipole-moment matrix elements. Given that 
accurate dipole calculations are challenging, this feature alone eliminates the most significant
source of numerical noise, and makes it possible to calculate HHG spectra with the dynamic range
well beyond what is typical for the traditional approach.

\section{HHG in zinc-blende materials}

To illustrate the capabilities of the SBE-solver algorithm described in the previous sections,
we present simulations of high-harmonic generation in zinc-blende structures, choosing
GaAs and ZnSe for our examples.

\smallskip
\noindent{\bf Material model}\\
As for the choice of the material model, most of the HHG simulations 
utilize  DFT calculations to obtain the material band-structure and related quantities such
as dipole moments. While the method described above is in principle applicable with any material description
capable of producing Hamiltonian eigenstates for any $\mathbf{k}$ throughout the Brillouin zone, for this
work we prefer to use tight-binding models. One could argue that such a description is less accurate
in terms of the band-structure, and it is a valid point. On the other hand, SBE-based simulations
using DFT-based material model over a three-dimensional Brillouin zone has yet to be demonstrated. 
Moreover, the results from DFT calculations suffer from numerical issues, for example it may be difficult
to tell apart Bloch states which are energetically close from truly degenerate states. Because we concentrate on
qualitative properties of the HHG, for this work we choose the tight-binding description which is free of such numerical issues.

We have used the empirical tight-binding models to obtain the quantities required by the solver,
i.e. the $\mathbf{k}$-dependent Hamiltonian $h(\mathbf{k})$ built on the frequently
used sp$^3$s$^*$ model~\cite{VOGL,TBM2}. For simplicity, we neglect the spin-orbit coupling, and include ten bands.
The explicit form of the Hamiltonian matrix and its parameterization can be found in Ref.~\cite{TBM1}.
Exact diagonalization procedure is executed ``on the fly'' as needed for any
given $\mathbf{k}$, producing the set of eigenvectors $\vert m \mathbf{k}\rangle$ and band energies
$\epsilon(\mathbf{k})$. For the calculation
of the current-density (\ref{eq:observeJ}), the vector matrix $\partial_{\mathbf{k}}h(\mathbf{k})$ is also calculated
exactly from the model. Thus, there is no interpolation or any finite-difference approximations needed.

\smallskip
\noindent{\bf Pulsed excitation }\\
The examples given next assume excitation by a linearly polarized pulse with the central wavelength of $\lambda=3.6\mu$m,
envelope duration of 100~fs ($\cos^2$ shape), and the field intensity of $8.7\times 10^8$V/m. We explore different
crystal orientations in order to demonstrate that the nonlinear response exhibits the expected orientation
and polarization properties. 
As propagation effects~\cite{Xia:18} are not studied in this work, the observable of interest is the vector of the current-density
calculated for the given excitation pulse.

\begin{figure}[t]
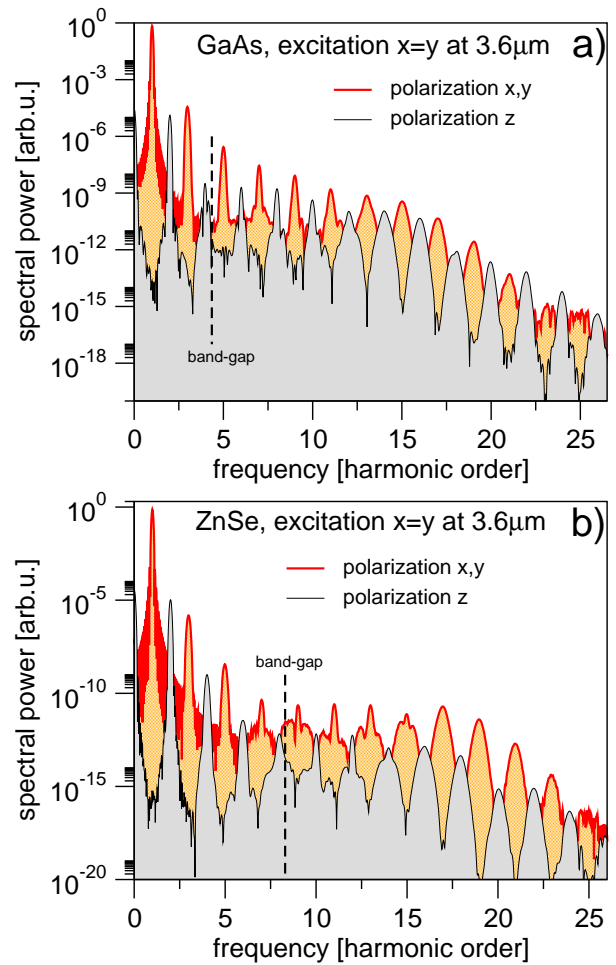

\centerline{\includegraphics[clip,width=0.92\linewidth]{./figure1a.eps}}
\centerline{\includegraphics[clip,width=0.92\linewidth]{./figure1b.eps}}
\caption{
  High-harmonic generation in  crystals excited by a linearly polarized
  pulse oscillating along the $x\!\!=\!\!y$  (crystal) direction. 
\label{fig:100spc45}
  }
\end{figure}

\smallskip
\noindent{\bf HHG-spectra from the whole Brillouin zone}\\
For the first example we consider a crystal sample oriented such that the linearly polarized pulse
oscillates along direction (1,1,0), i.e. perpendicular to the crystal $z$-axis. In this geometry,
the material symmetry dictates that the second-harmonic response only appears in the $z$-direction.
This is because the second-order tensor $\chi_{abc}^{(2)}$ of the zinc-blende structure vanishes unless
all $a,b,c$ are different. In contrast, the third harmonic excited by the Kerr effect  is expected to
show up along the $x\!=\!y$ direction.

Figure~\ref{fig:100spc45} depicts the simulated HHG spectra for GaAs and ZnSe samples, and shows that the
polarization properties are indeed as one expects, with even and odd harmonics are separated
between the parallel and perpendicular polarizations.

We have intentionally used a relatively long-duration pulse so that the well-separated  harmonics 
showcase that the calculated spectra are free of the noise-floor so typical of many HHG simulations ---
here the noisy background occurs 
about ten orders of magnitude below the lower edge of these plots. This indicates
excellent numerical fidelity of the algorithm.

\begin{figure}[t]
  \centerline{\includegraphics[clip,width=0.92\linewidth]{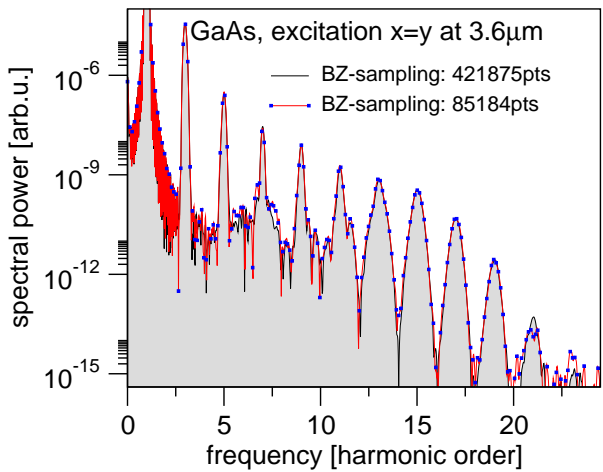}}
\caption{
  Convergence of the numerical HHG-spectra for two different numbers
  of sampling points in the 3D Brillouin zone.
\label{fig:spcconv}
  }
\end{figure}

The question of convergence is obviously important. Making sure that the time-step is
short enough and does not affect the convergence, we compare the spectra simulated with
different number of sampling points in the Brillouin zone. Figure~\ref{fig:spcconv}
shows an example where convergence is achieved over a dynamic range of fifteen
orders of magnitude.

\begin{figure}[t]
    \centerline{\includegraphics[clip,width=0.92\linewidth]{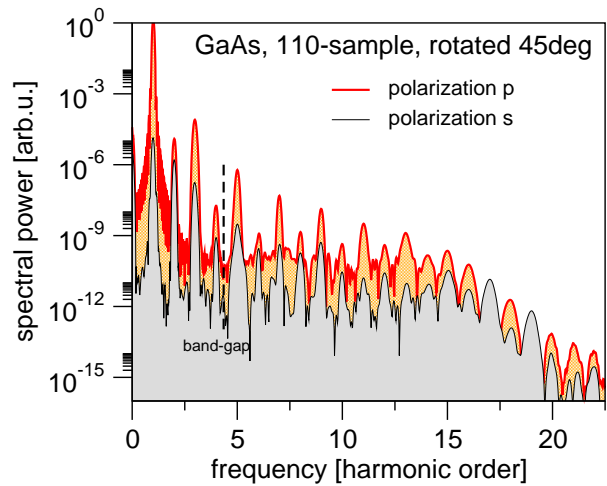}}
\caption{
  High-harmonic generation in 110-oriented GaAs crystal. Linearly polarized
  excitation pulse oscillates at 45 degrees w.r.t. the crystal axis.
  In this geometry, both even and odd harmonics should appear in  the
  $p$- as well as in the $s$-polarization.
\label{fig:110spc45}
  }
\end{figure}

To show a case when both even and odd harmonics appear simultaneously in the
parallel and perpendicular polarizations, we include Fig.~\ref{fig:110spc45}.
Although we do not actually propagate the excitation pulse, we assume
that the sample orientation is 110, and then rotate the sample about the beam axis
as it is often done in experiments. In this figure the sample is rotated by 45 degrees,
and we look at the current density polarized parallel (p) and perpendicular (s)
to the polarization direction of the excitation pulse. In this particular case,
the even harmonics, while clean and well defined, are weaker than the odd harmonics
and this is especially the case for the p-polarized component.

\begin{figure}[t]
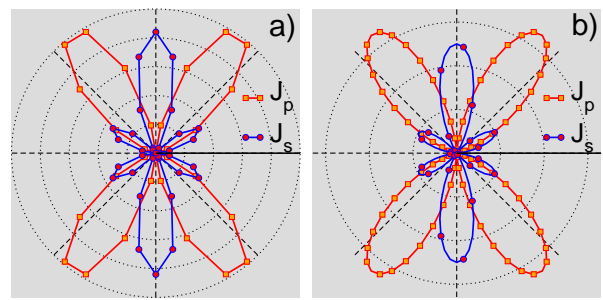

  \centerline{\includegraphics[clip,width=0.45\linewidth]{./figure4a.eps}
              \includegraphics[clip,width=0.45\linewidth]{./figure4b.eps}}

  \caption{
  Orientation-dependent high-harmonic generation in a 110-oriented GaAs crystal.
  Left: Simulated second harmonic filtered from the current density
  in the parallel (p) and perpendicular (s) polarizations
  shown as functions of the sample rotation angle. Right: The second harmonic
  radiation pattern  for the zinc-blende nonlinear tensor.
\label{fig:110spcANG}
  }
\end{figure}

The relative strength between the odd and even harmonics depends on the
angle of the sample rotation.
This is illustrated in Fig.~\ref{fig:110spcANG} for the
second-harmonic frequency band. The radiation pattern (left) is essentially the same as
expected from the classical $\chi^{(2)}$ tensor of the zinc-blende
structure (right), and this corroborates that the simulated response has the correct
symmetry. It should be interesting to study the deviations from the
classical (equilibrium) predictions based on a fixed  $\chi^{(2)}$
as a function of the excitation pulse intensity, but we will not
pursue this here.

\begin{figure}[t]
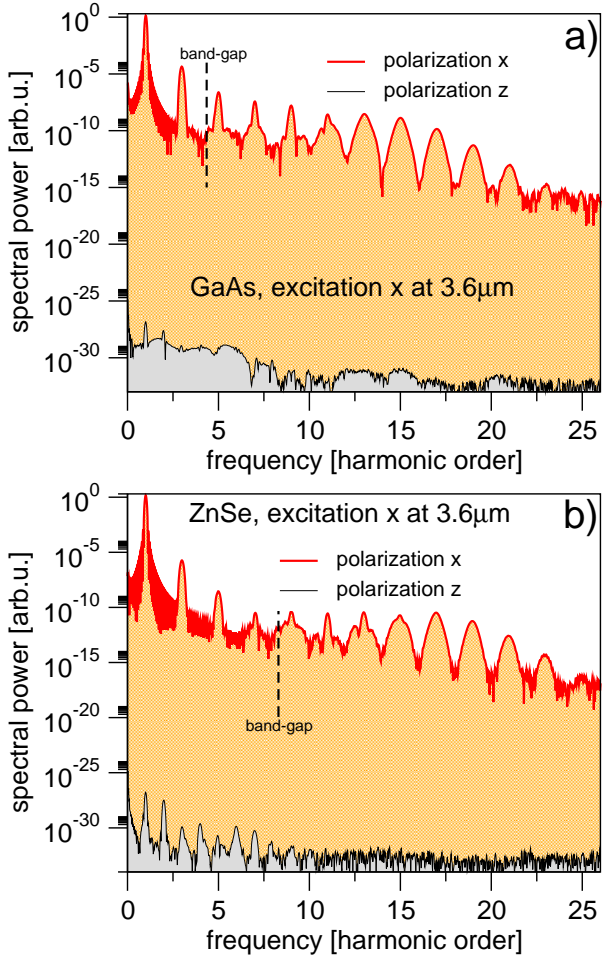

\centerline{\includegraphics[clip,width=0.92\linewidth]{./figure5a.eps}}
\centerline{\includegraphics[clip,width=0.92\linewidth]{./figure5b.eps}}
\caption{
  High-harmonic generation in  crystals excited by a linearly polarized
  pulse oscillating along the (crystal) $x$ direction. In this geometry, the current
  density in the perpendicular direction ($y,z$) must vanish due to the
  material symmetry. 
\label{fig:100spc0}
  }
\end{figure}

For a more difficult-to-pass test of the symmetry properties of the simulated
high-harmonics, Fig.~\ref{fig:100spc0} shows the results for the excitation
with a pulse polarized along one of the crystal axes ($x$). In this case the
response components $z$ and $y$ are supposed to vanish and they indeed do.
The $z$-component shows up in these plots as a noisy background (gray area below the
black curve) about fifteen orders of magnitude below the level of the  $x$-polarized
signal. One could say that this is nothing but a simple sanity check because
our SBE-based simulation automatically inherits the correct symmetry
properties from the material model. Nevertheless, it is important to note that
the ``numerical zero'' demonstrated  for the current components
which are forbidden by symmetry does not occur  point by point (in the reciprocal space).
Instead, all regions throughout the Brillouin zone contribute non-zero signals,
and the symmetry appears only after significant (or complete in the case here) cancellations.
Because of their important implications, we discuss these issues next.

\smallskip
\noindent{\bf Mapping the Brillouin zone for the HHG-source}

One often utilized simplification in the solid-state HHG simulations
is that instead of the entire Brillouin zone only a one dimensional
line is used to represent the reciprocal space. We now present a few examples
which demonstrate that a great deal of caution is in order when trying
to interpret HHG-simulation results based on a low-dimensional subset of
the reciprocal space because:\\
a) the source of the high-harmonics is distributed throughout the entirety
of the Brillouin zone\\
b) different portions of the zone give rise to radiation with various
phase shift and significant cancellation occur between them.\\
The illustrations also elucidate how it happens that the second-harmonic
signals are absent in Fig.~\ref{fig:100spc0}

\begin{figure}[t!]
  \centerline{\includegraphics[clip,width=0.3 \textwidth]{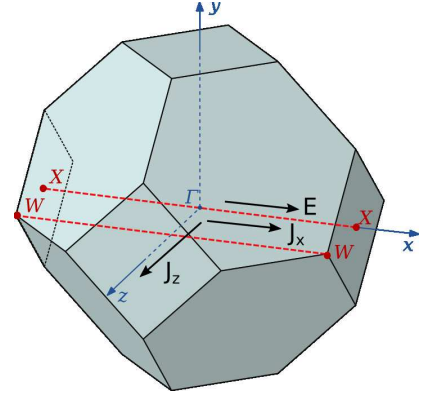}}
\caption{
  \label{fig:BZlines}
  Lineouts through the first Brillouin zone are shown in dashed red lines.
  The X$\Gamma$X path is often used to simulate HHG from materials such as GaAs.
  We wish to visualize the contribution of different points along the lineout,
  and see how they change when moving away from the center of the zone to, say,
  the WW-lineout. The arrows indicate the direction of the excitation ($E$) and induced
  current components ($J_x,J_z$) corresponding to the components shown in Fig.\ref{fig:100spc0}.
  }
\end{figure}

Let us consider a lineout of the Brillouin zone, for example a line of
$\mathbf{k}$-vectors connecting two W points at the opposite sides of the Brillouin zone,
or the  X-$\Gamma$-X path going through the center of the zone, as
depicted in Fig.\ref{fig:BZlines}. Black arrows indicate the polarization direction
of the electric field, and of the parallel ($J_x$) and perpendicular ($J_z$)
component of the induced current.

We calculate $\rho(\mathbf{k};t)$ for each point of such a lineout and
evaluate the corresponding current density as the trace with $\partial_{\mathbf{k}} h(\mathbf{k})$
as required by (\ref{eq:observeJ}).
The result is a contribution to the current which originates in the
electronic states starting their evolution at a point of the lineout. 
We aim to compare the ``strength of the response'' between different
regions of the reciprocal space.

\begin{figure}[t!]
\centerline{\includegraphics[clip,width=0.4 \textwidth]{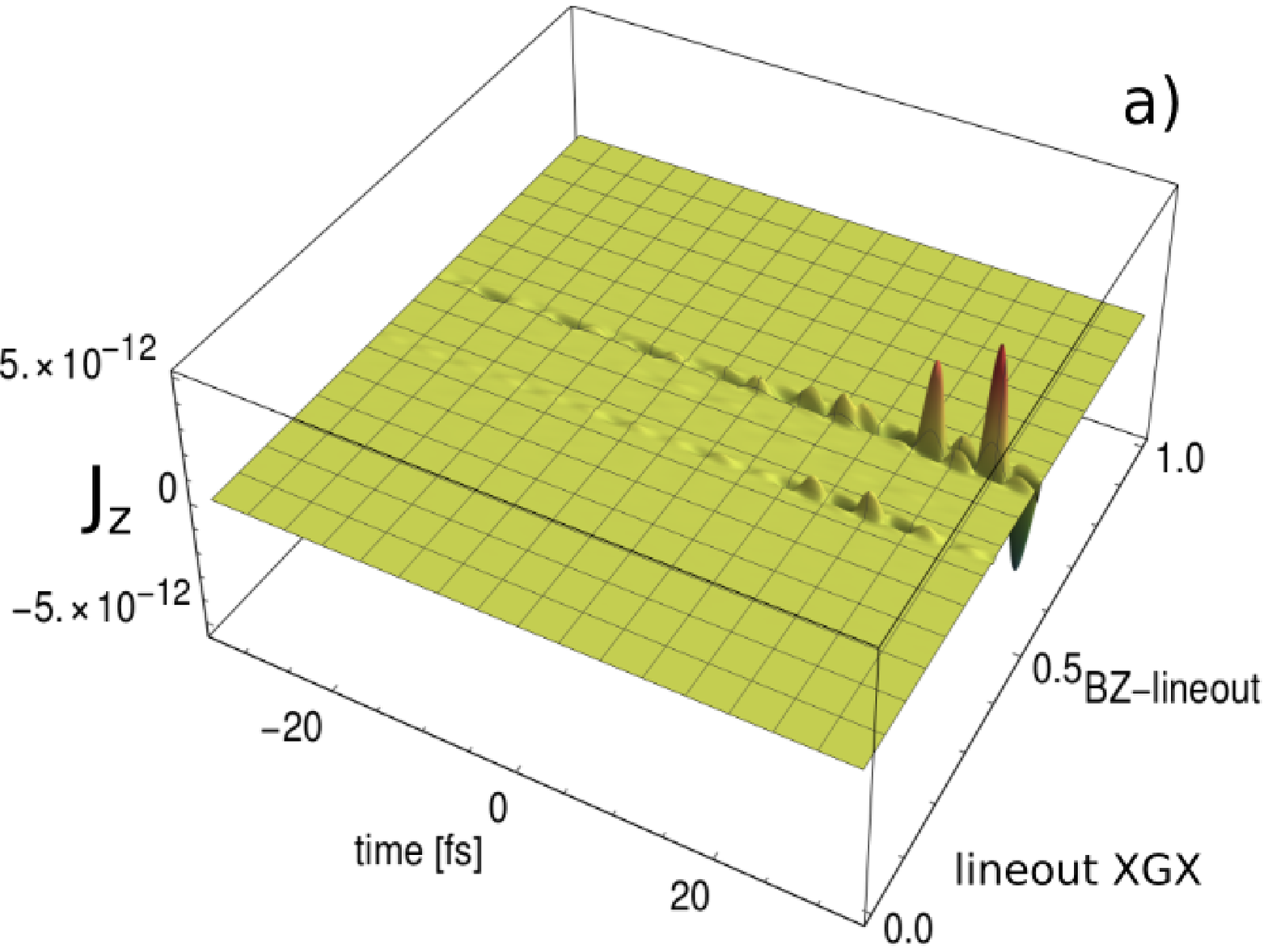}}
\centerline{\includegraphics[clip,width=0.4  \textwidth]{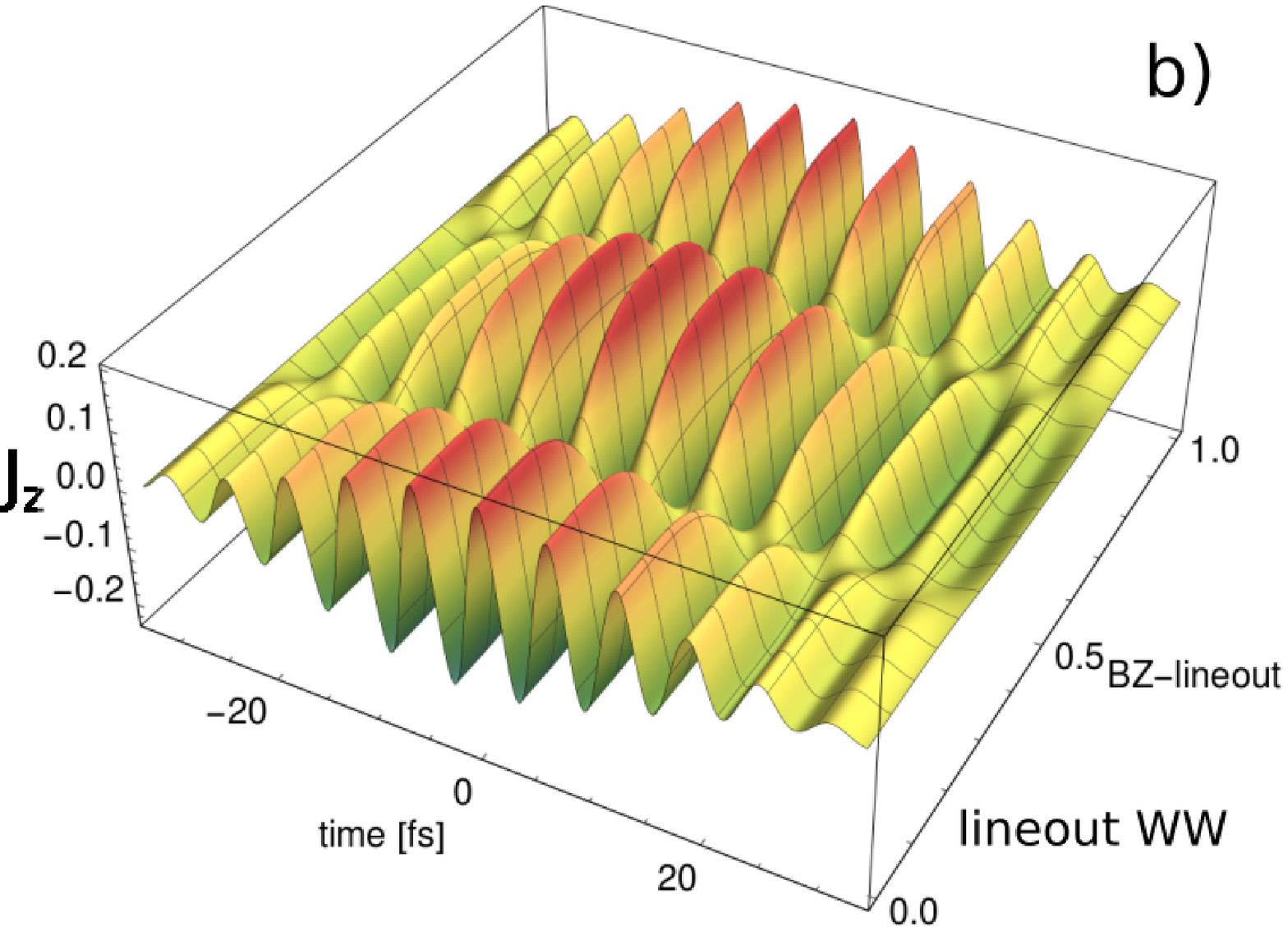}}
\caption{
  \label{fig:lineout1}
  The vector component of the time-dependent current density  perpendicular to the
  driving electric field. Shown here is the filtered second-harmonics in arbitrary
  units as a function of the initial $\mathbf{k}$-vector
  localized along the indicated lineout (dashed red line in Fig.\ref{fig:BZlines}) of the Brillouin zone.
  The lineout axis is in relative units.
  }
\end{figure}

Instead of the HHG-spectrum, we visualize the induced current-density 
because in this way one can appreciate different phase shifts
and see how various contributions can interfere.
In order to make figures easier to read we assume a shorter pulse, 50~fs
duration, and we filter out the second-harmonic contribution from the current-density.
Then we plot a two-dimensional map of the current  versus
time and the initial $\mathbf{k}$-location along the selected lineout.

To elucidate the mechanism behind the vanishing second harmonics in Fig.~\ref{fig:100spc0}.
we first  consider the z-polarization output shown in Fig.~\ref{fig:lineout1}
for the lineouts X$\Gamma$X (top) and WW (bottom). What the top  plot shows
is merely numerical noise, so we can see that the points along  X$\Gamma$X
do not generate the s-polarized SH contributions at all. However, moving away from the
axis of the Brillouin zone to the line WW (bottom panel),  one can see that every point gives a strong
individual contribution, and it is because the middle and outer portions of
the lineout are out of phase  that the total second harmonic  vanishes in the end.

\begin{figure}[t]
\centerline{\includegraphics[clip,width=0.4 \textwidth]{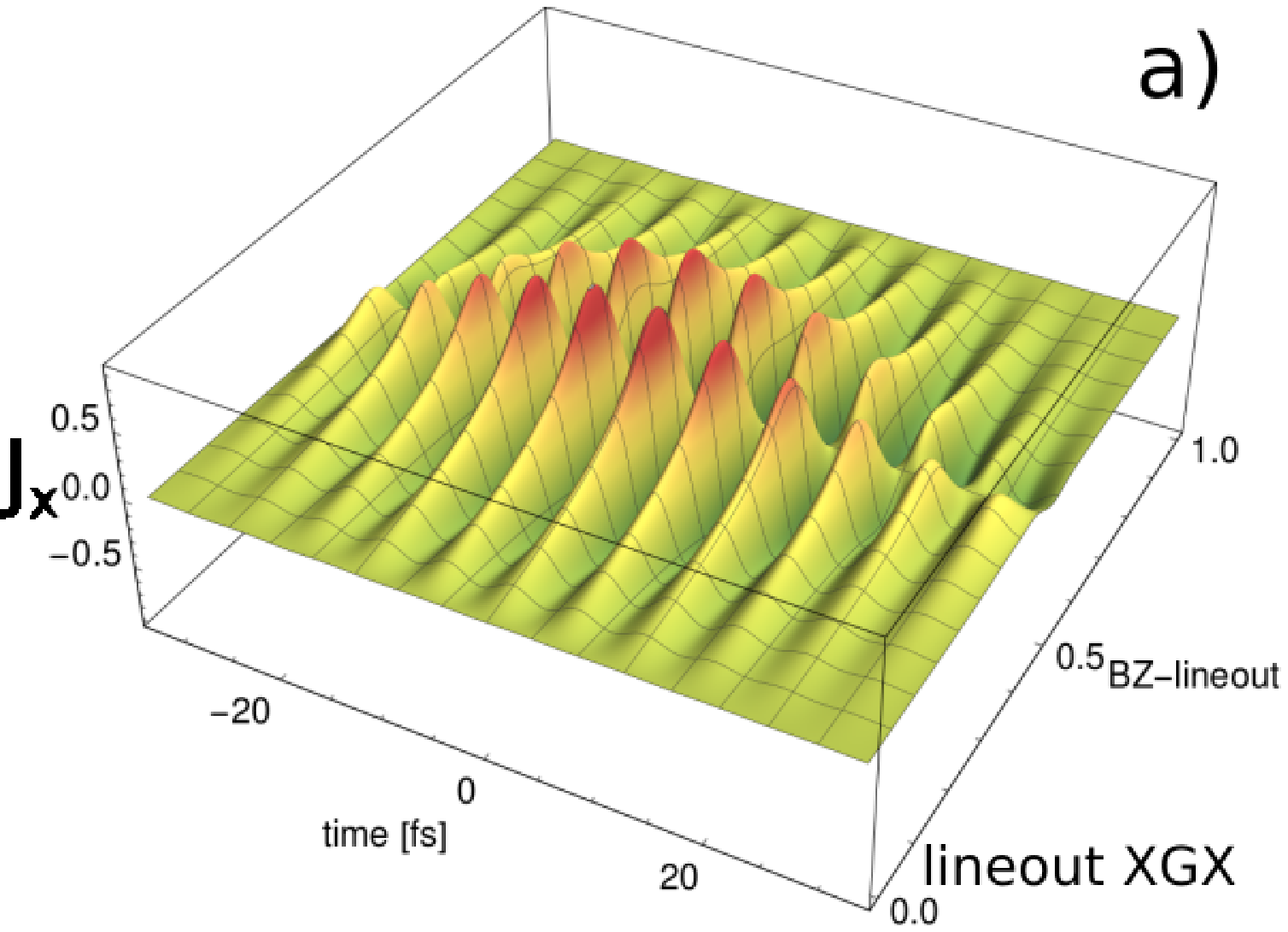}}
\centerline{\includegraphics[clip,width=0.4 \textwidth]{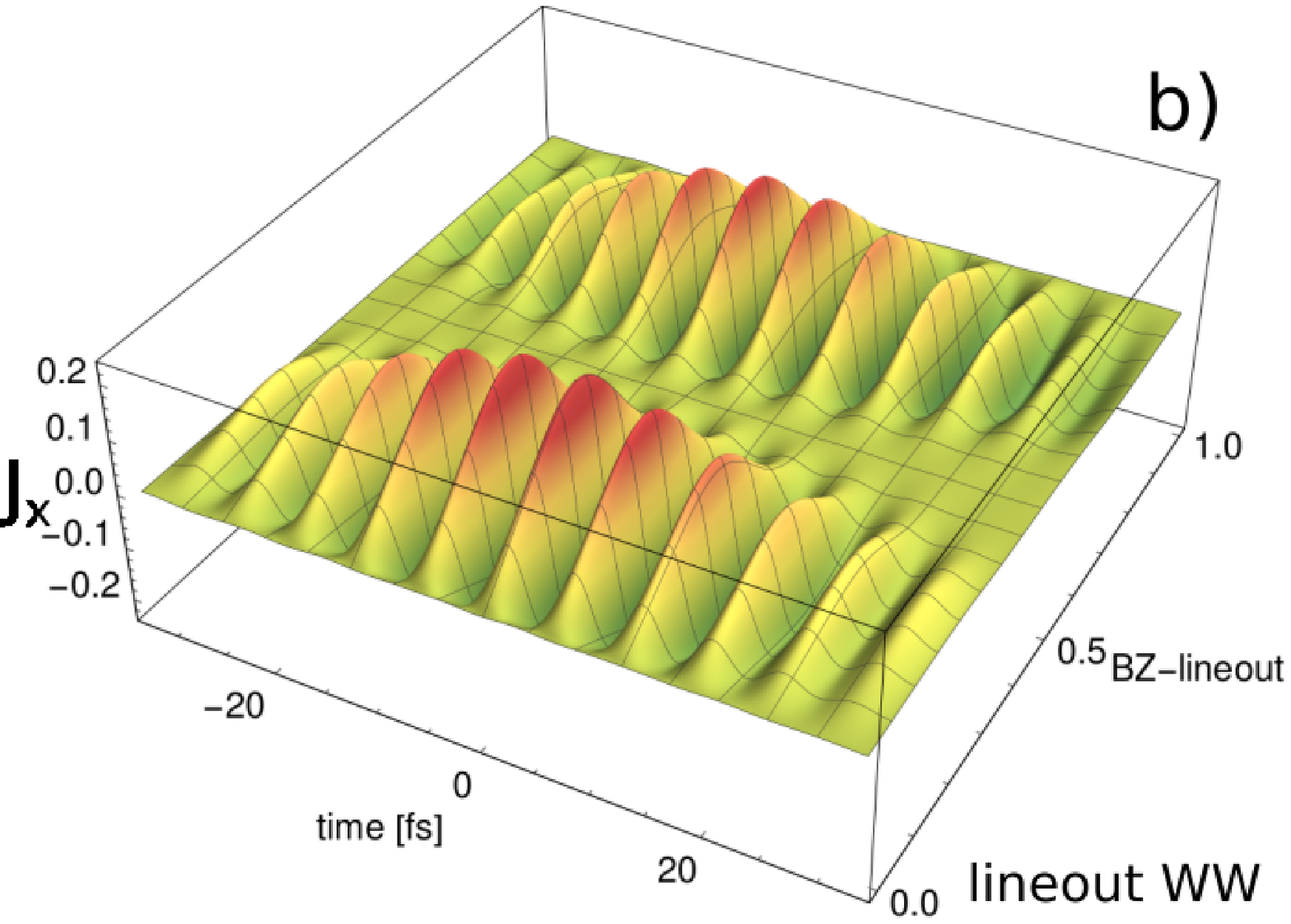}}
\caption{
\label{fig:lineout2}
Time-dependent current density as in Fig.\ref{fig:lineout1}, but for the
current-density vector component parallel with the electric field. 
}
\end{figure}

The mechanism that extinguishes the second harmonic for the polarization along the electric field direction
(cf. absent second-harmonic peak in the red (top) lines in Fig.~{\ref{fig:100spc0})
is similar and is illustrated in Fig.~\ref{fig:lineout2}. This time we see strong contribution
along both lineouts, but different regions in the reciprocal space exhibit
out-of-phase contribution that interfere destructively.

These results are merely examples which of course can not provide a complete ``map'' of how different
parts of the Brillouin zone contribute to the observed HHG. Nevertheless, they make it quite evident that
all parts of the Brillouin zone contribute to the HHG output on a qualitatively equal footing, and
only when they are added together the correct picture emerges. It is obvious that for a sample rotated
with respect to that in the above example, the resulting strength of the harmonics of different polarization
will sensitively reflect the interference between different parts of the Brillouin zone.

Our results
also suggest that it is not given that the observed response is dominated by the initial $\mathbf{k}$-states
with the strongest transition dipoles. Indeed, the $\mathbf{k}$-dependence of the signal-amplitudes in
Figs.~\ref{fig:lineout1} and \ref{fig:lineout2} does not follow the magnitude of the dipole moments which tend to
be strongest in the vicinity of the $\Gamma$-point.
To emphasize this even more, 
Fig.~\ref{fig:lineout3} shows an example for a
crystal sample with 110 orientation rotated about the beam by 90 (top panel) and 45 (bottom panel) degrees.
While in this case the response from different $\mathbf{k}$ location appears to
be in phase, the bottom panel shows that the strongest response depends on the polarization; when in the
upper panel it is correlated with the strongest dipoles in the center, and the lower panel exhibits
an asymmetry  which is ``out of sync'' with the magnitude of the local transition dipoles. Moreover,
it becomes evident that this particular BZ-lineout should not be considered in isolation
from its counterparts related by the crystal symmetry.

\begin{figure}[t]
  \centerline{\includegraphics[clip,width=0.4 \textwidth]{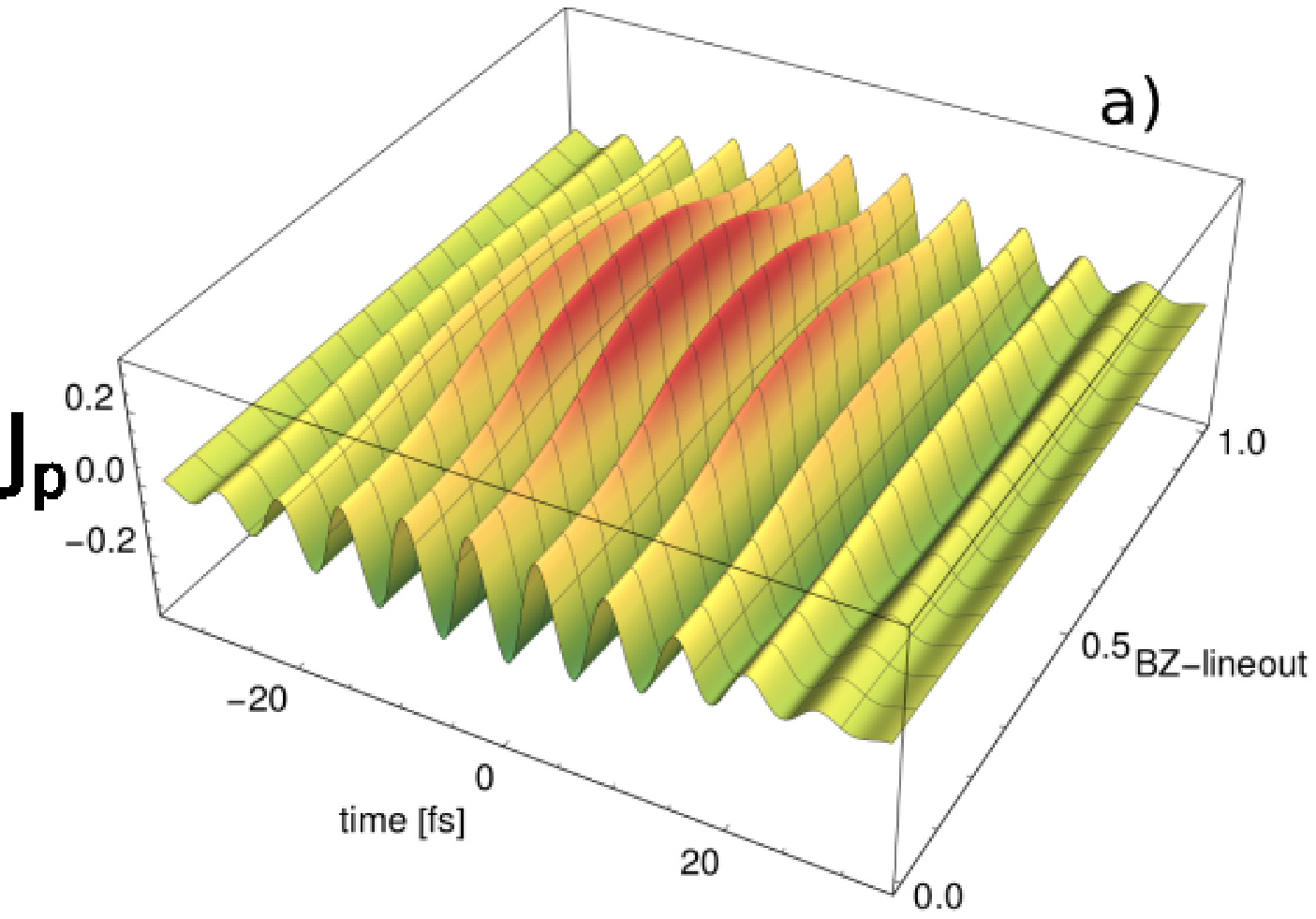}}
  \centerline{\includegraphics[clip,width=0.4 \textwidth]{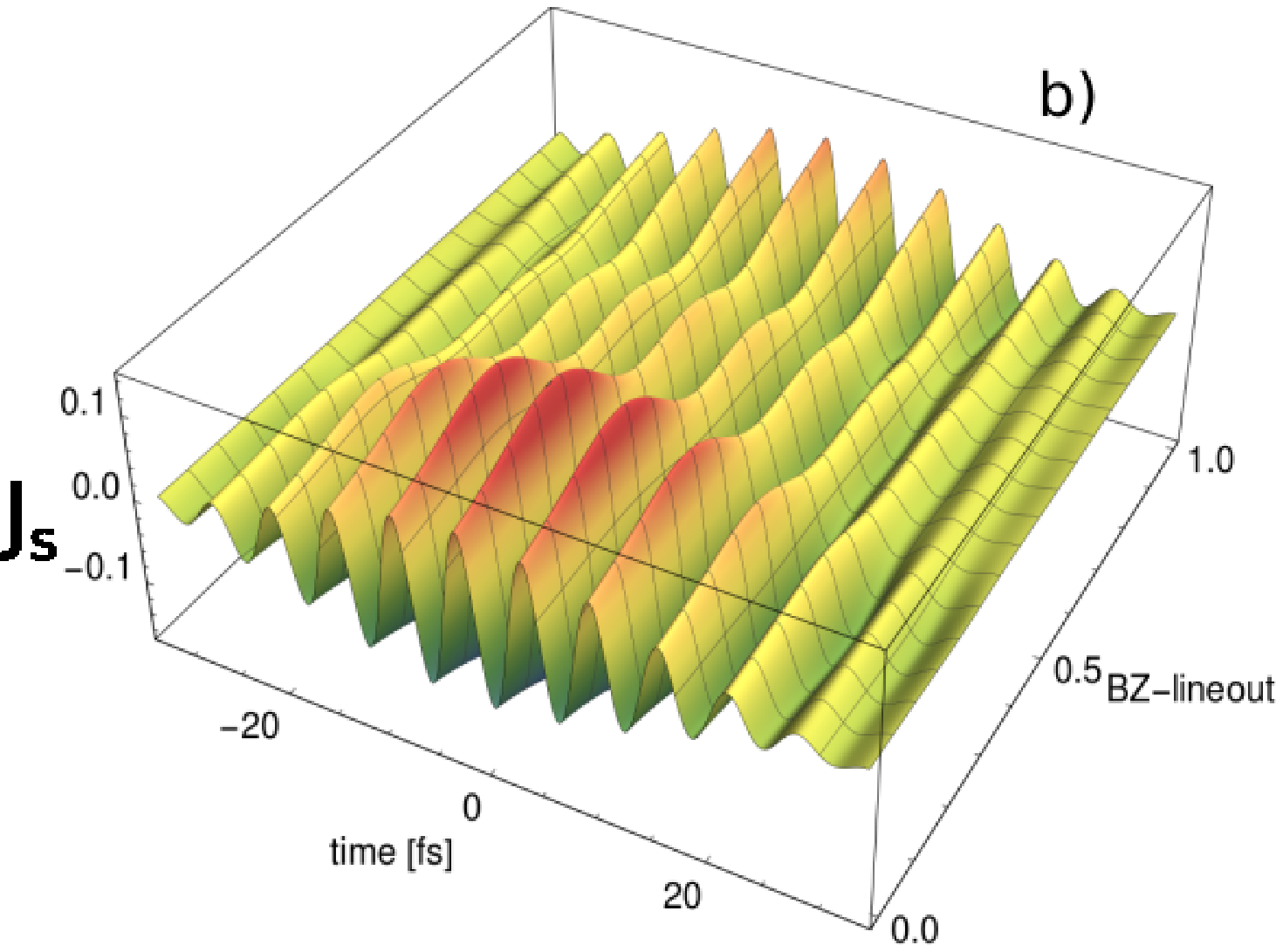}}
\caption{
\label{fig:lineout3}
Time-dependent current density (in arbitrary units) for the WW lineout
(red dashed line in Fig.\ref{fig:BZlines}) and 110-oriented
crystal sample. The top panel shows the p-polarized response for the sample rotated by 90 degrees,
while the lower panel is for the s-polarized response and sample rotated by 45 degrees.
It is obvious that the induced current amplitude is not always correlated with the local magnitude
of the transition dipoles which are strongest in the middle of the lineout.
}
\end{figure}

We therefore contend that the integration over the entirety of the Brillouin zone should  be the default approach
preferred over the numerically less intensive investigations restricted to low-dimensional subsets in the reciprocal space.

\section{Conclusions}

We have presented an approach to the high-harmonic generation from crystalline solid-state media
which is completely free of any considerations related to the complex phases of the elements
of the transition-dipole moment. In fact, the method does not require calculations of the
transition dipole matrices which is a distinct advantage by itself. The simulation algorithm is informed
by the fact that the absolute phases of these quantities are not physical observables, and the method is 
``phase-blind'' by design in the sense that it can work with arbitrary phases assigned 
to the Hamiltonian eigen-states. In particular, there is no requirement of differentiability or
even continuity between the Hamiltonian bases used at ``mutually close'' points of the Brillouin zone.
As such, our approach offers
the best possible solution to the so-called transition-dipole phase problem by eliminating the issue entirely.

The method is computationally efficient and admits a perfectly load-balanced parallelization.
The speed is sufficient for future integration with the pulse-propagation simulators such as our gUPPE~\cite{gUPPE},
making the spatially resolved studies of propagation effects in solid-state HHG feasible with the account
of the whole Brillouin zone.

It is actually relatively
easy to integrate all induced current-density contributions over the whole three-dimensional Brillouin zone. This
is shown crucial for the preservation of the material symmetry. Once the initial model utilized to
calculate the band-structure of the crystalline medium properly reflects the space-group of the
material, the simulated HHG signals and in particular their sample-orientation and pulse-polarization dependencies
are guaranteed to be correct.

We have shown that in general the entire Brillouin zone contributes to the high-harmonic signal.
This is perhaps not so surprising, but our simulation examples also demonstrate that
there are considerable cancellations, or destructive interference between the contributions originating
from the quantum states in different sectors of the Brillouin zone. It is therefore unrealistic to expect that,
for a general sample orientation and excitation-pulse polarization, one could use a low-dimensional
subset of the Brillouin zone to capture the high-harmonic generation very accurately.
We have also seen that the strength of the transition-dipoles is not a reliable predictor of
which part of the Brillouin zone may dominate the HHG signal.

These observation may have important impact on some applications of solid-state HHG, such as Berry curvature
measurement~\cite{BerryCurvature}. For example,
all-optical band-reconstruction~\cite{ReconstAllOptBands,ReconstBands} and dipole-reconstruction~\cite{ReconstDipole}
methods tend to utilize a one-dimensional picture
of the reciprocal space by  selecting a presumably dominant contribution to the process~\cite{MapAssume0} 
in order to extract information concerning the material band-structure.
In contrast, here we have seen how the HHG-source can ``light up'' the Brillouin zone in rather non-intuitive
patterns. We therefore believe that the full-Brillouin zone simulations similar to those presented in our
work can be a useful tool to identify the dominant channels in the high-harmonic generation from crystalline materials.

\medskip

\centerline{\bf ACKNOWLEDGMENTS}
This research was supported by the US Army Research Laboratory under grant no.
W911NF1920192, and by the
Air Force Office for Scientific Research under grants no.
FA9550-22-1-0182
and
FA9550-21-1-0463.


%

\end{document}